\documentclass[twocolumn,aps,showpacs,preprintnumbers, superscriptaddress,prl]{revtex4-1}
\usepackage{graphicx}
\usepackage{dcolumn}
\usepackage{amssymb}
\usepackage{bm}
\usepackage[utf8]{inputenc}
\usepackage{amsmath,amsfonts,amssymb}
\usepackage{color}
\usepackage[sf]{subfigure}
\usepackage{fouriernc}
\usepackage{xcolor}

\usepackage[colorlinks=true,
            linkcolor=blue,
            urlcolor=blue,
            citecolor=blue,
            unicode,
            pdfencoding=auto]{hyperref}
\usepackage{etoolbox}

\usepackage{soul}

\begin{document}

\title{Chiral Gravitons in Fractional Quantum Hall Liquids}

\author{Shiuan-Fan Liou}
\affiliation{NHMFL and Department of Physics, Florida State University, Tallahassee, Florida 32306, USA}

\author{F. D. M. Haldane}
\affiliation{Physics Department, Princeton University, Princeton, New Jersey 08544, USA}

\author{Kun Yang}
\affiliation{NHMFL and Department of Physics, Florida State University, Tallahassee, Florida 32306, USA}

\author{E. H. Rezayi}
\email{erezayi@calstatela.edu}
\affiliation{Physics Department, California State University Los Angeles, Los Angeles, California 90032, USA}

\pacs{73.43.Nq, 73.43.-f}

\begin{abstract}
We elucidate the nature of neutral collective excitations of fractional quantum Hall
liquids in the long-wavelength limit. We demonstrate that they are chiral gravitons carrying angular momentum -2, which are quanta of quantum motion of an
internal metric, and show up as resonance peaks in the systems response to what is the fractional Hall analog of gravitational waves. Relation with existing and possible future experimental work that can 
detect these fractional quantum Hall gravitons and reveal their chirality are discussed.
\end{abstract}

\date{\today}

\maketitle

{\em Introduction and Motivation} --
Study of neutral excitation spectra of fractional quantum Hall (FQH) liquids has a long history.
It is now well-understood that there is a sharp magneto-roton mode exhibiting a ``roton" minimum at a finite wave vector\cite{GMP}, which has been observed
experimentally\cite{klitzing}. The observability of this mode is related to the fact that it is a bound state of a Laughlin quasiparticle-quasihole pair (or
an quasi-exciton), which can be created by the (Landau-level projected) density operator, which is dipole-active; it can thus be excited electromagnetically at the
appropriate {\em finite} wave vector.
The sharpness of the magneto-roton mode is tied to the fact that its energy is below the continuum formed by more complicated multi-quasiparticle/quasihole
excitations (or multi-rotons).

On the other hand the understanding of collective excitations at long-wavelengths  is far from complete. The magneto-roton dispersion enters the continuum as the
wave vector decreases, making it very hard to identify, and even casting doubt on its presence. More serious is the fact that Kohn's theorem dictates that dipole
spectral weight is exhausted by the cyclotron mode, making the single mode approximation\cite{GMP} completely ineffective at zero wave vector. This also renders any
long-wavelength intra-Landau level collective excitation invisible to electromagnetic probe in the linear response regime.

In a parallel line of work, one of us\cite{Metric} pointed
out that there exists an internal geometrical degree
of freedom (or internal metric) responsible for the intra-Landau level dynamics of the system, that is not properly captured by the standard description of FQH
liquids in terms of topological quantum field theories. Physical implications of this geometrical degree
of freedom has been discussed extensively\cite{qiu,haowang,Apalkov,boyang,Johri}, in particular its experimental
observability\cite{yang,kamburov,jo,Ippoliti,Ippoliti1,Ippoliti2}. Furthermore, it has also been argued\cite{Metric,boyang12,xiluo,golkar} that this internal metric
has its own quantum
dynamics, which gives rise to the long-wavelength collective excitations in FQH liquids that can be viewed as ``gravitons". This provides a new insight into the
invisibility of long-wave length intra-Landau level collective mode to electromagnetic probes: the graviton carries total angular momentum 2, mismatching that of
the photon which carries angular momentum 1.
In a recent paper, another of us\cite{yang16} argued that these ``gravitons" can instead be excited and probed using acoustic waves in the crystal, whose effects
mimic those of gravitational waves.

In this paper we present numerical results that demonstrate unequivocally the presence of the graviton mode, which shows up as a pronounced peak in the spectral
function of the dynamical gravitational response\cite{yang16}. We further reveal the chiral nature of the gravitons, namely they come with a specific polarization
corresponding to angular momentum -2.
We will discuss possible experimental probes of these gravitons and in particular, their polarization, as well as the relation between our results and closely related works.

{\em Spectral Functions} -- Ref. \cite{yang16} considered the coupling between an oscillating effective mass tensor and the intra-Landau level degrees of
freedom of a two-dimensional electron gas (2DEG) confined to the lowest Landau level (LLL). For a two-body interaction of the form
\begin{eqnarray}
V^{(2)}=\sum_{i < j} V({\bf r}_i-{\bf r}_j)=\frac{1}{2}\sum_{\bf q}V_{\bf q}\rho_{\bf q}\rho_{-{\bf q}},
\end{eqnarray}
where $V_{\bf q}$ is the Fourier transform of electron-electron interaction potential $V({\bf r})$ and
\begin{eqnarray}
\rho_{\bf q}=\sum_{i}e^{i{\bf q}\cdot{\bf r}_i}
\end{eqnarray}
is the density operator, it was found that the coupling is described by the operator
\begin{eqnarray}
\hat{O}^{(2)}=\sum_{\bf q}(q^2_y - q^2_x)V_{\bf q}e^{-\frac{1}{2}q^2\ell^2}\overline{\rho}_{\bf
q}\overline{\rho}_{-{\bf q}},
\label{eq:gravitycoupling}
\end{eqnarray}
in which
\begin{eqnarray}
\overline{\rho}_{\bf q}=\sum_{i}e^{i{\bf q}\cdot{\bf R}_i}
\end{eqnarray}
is the LLL projected density operator, and ${\bf R}$ is the guiding center coordinate.
It is straightforward to generalize the above to multi-particle interactions; of particular relevance to our later discussion is the case of a three-body
interaction
\begin{eqnarray}
V^{(3)}=\frac{1}{6}\sum_{{\bf q}_1{\bf q}_2}V_{{\bf q}_1{\bf q}_2}\rho_{{\bf q}_1}\rho_{{\bf q}_2}\rho_{-{\bf q}_1-{\bf q}_2},
\end{eqnarray}
in which case the coupling is given by
\begin{eqnarray}
\hat{O}^{(3)}&=&\sum_{{\bf q}_1{\bf q}_2}V_{{\bf q}_1{\bf q}_2}(q^2_{1y}+q^2_{2y}+q_{1y}q_{2y}  - q^2_{1x} - q^2_{2x}- q_{1x}q_{2x})\nonumber\\
&\times& e^{-\frac{\ell^2}{2}(q_1^2+q_2^2+|{\bf q}_1+{\bf q}_2|^2)}\overline{\rho}_{{\bf q}_1}\overline{\rho}_{{\bf q}_2}\overline{\rho}_{-{\bf q}_1-{\bf q}_2}.
\label{eq:gravitycoupling3}
\end{eqnarray}

Our numerical studies are based on the calculation of spectral functions of $\hat{O}$ and its close relatives $\hat{O}_\sigma$ (to be specified below) in finite-size systems:
\begin{eqnarray}
I_\sigma(\omega)=\sum_n|\langle n|\hat{O} _{\sigma}|0\rangle|^2\delta(\omega-\omega_n),
\label{eq:SpectralFunction}
\end{eqnarray}
where $|0\rangle$ is the ground state, $|n\rangle$ is an excited state with excitation energy $\hbar\omega_n$. (from here on we set $\hbar=1$.) This is the system's transition rate due to an
oscillating effective mass tensor metric, which is analogous to an oscillating metric induced by a gravitational wave.

To establish the chiral nature of the graviton, it is convenient to study operators that have ``handedness''
instead of the ones given in Eqs. (\ref{eq:gravitycoupling}) and (\ref{eq:gravitycoupling3}).  Therefore we define:
\begin{eqnarray}
\hat{O}_\mp^{(2)}=\sum_{\bf q}(q_x \mp i q_y)^2 V_{\bf q}e^{-\frac{1}{2}q^2\ell^2}\overline{\rho}_{\bf
q}\overline{\rho}_{-{\bf q}},
\label{eq:gravitycoupling_new}
\end{eqnarray}
where we have discarded an overall minus sign. A similar extension can be done for the 3-body or for that
matter to the n-body case of the Read-Rezayi\cite{RR} sequence. In this paper we will focus on the
Moore-Read (MR) state\cite{MR} with 3-body interactions.
As we will see $\hat{O}_\mp$ are the creation and annihilation operators of the graviton respectively, while $\hat{O}=(\hat{O} _{+} + \hat{O} _{-})/2$ is equivalent to the displacement operator in a harmonic oscillator, which couples to a linearly polarized ``gravitational wave"\cite{yang16}.

{\em Numerical Results} --
To compare  $I_\sigma(\omega)$ for different sizes it is convenient to normalize it, by dividing
out the factor $\langle 0 \lvert \hat{O}_\sigma^{\dagger} \hat{O}_\sigma \rvert 0 \rangle$, so that
\begin{eqnarray}
\int_{-\infty}^{\infty} I_\sigma(\omega) d\omega \ = \ 1.
\end{eqnarray}

\begin{figure}[t]
\centering%
\includegraphics[width=0.50\textwidth]{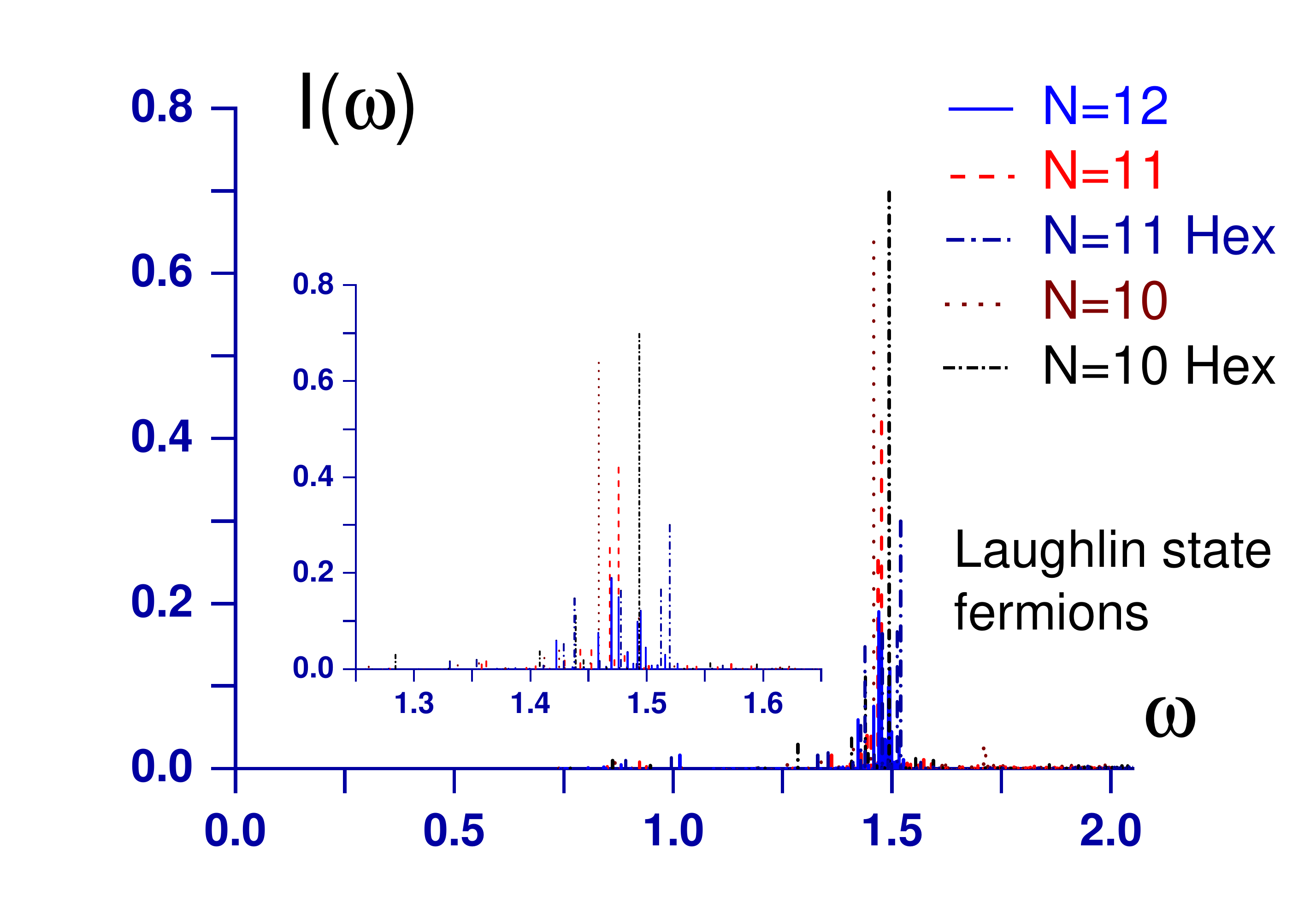}
\caption{\label{fig:LJFermi}%
A bird's-eye view of the $I(\omega)$ from fermionic Laughlin ground state at $\nu=1/3$ for various sizes and geometries (Hex stands for hexagon geometry; square geometry otherwise). The graviton response
can clearly be seen.  The inset shows more details near energies where the response is strong. For $N=10$ and 11 we recovered over 97\% of the total weights. Unlike
the case of bosons (shown below) the background noise seems to be stronger for fermions, which are also computationally more costly than bosons. Accordingly, for
$N=12$ we
recovered 90\% of the weight over 87\% of which is in the window of the inset. The rest appear to be background noise. For example, we recover over 99\% of the
weight for 10 electrons. However, no significant peaks other than those shown in the inset was seen. The total weight in the inset is only 87\% of the total weight.
In other words about 12\% is just background noise.
}
\end{figure}

\begin{figure}[t]
\centering%
\includegraphics[width=0.50\textwidth]{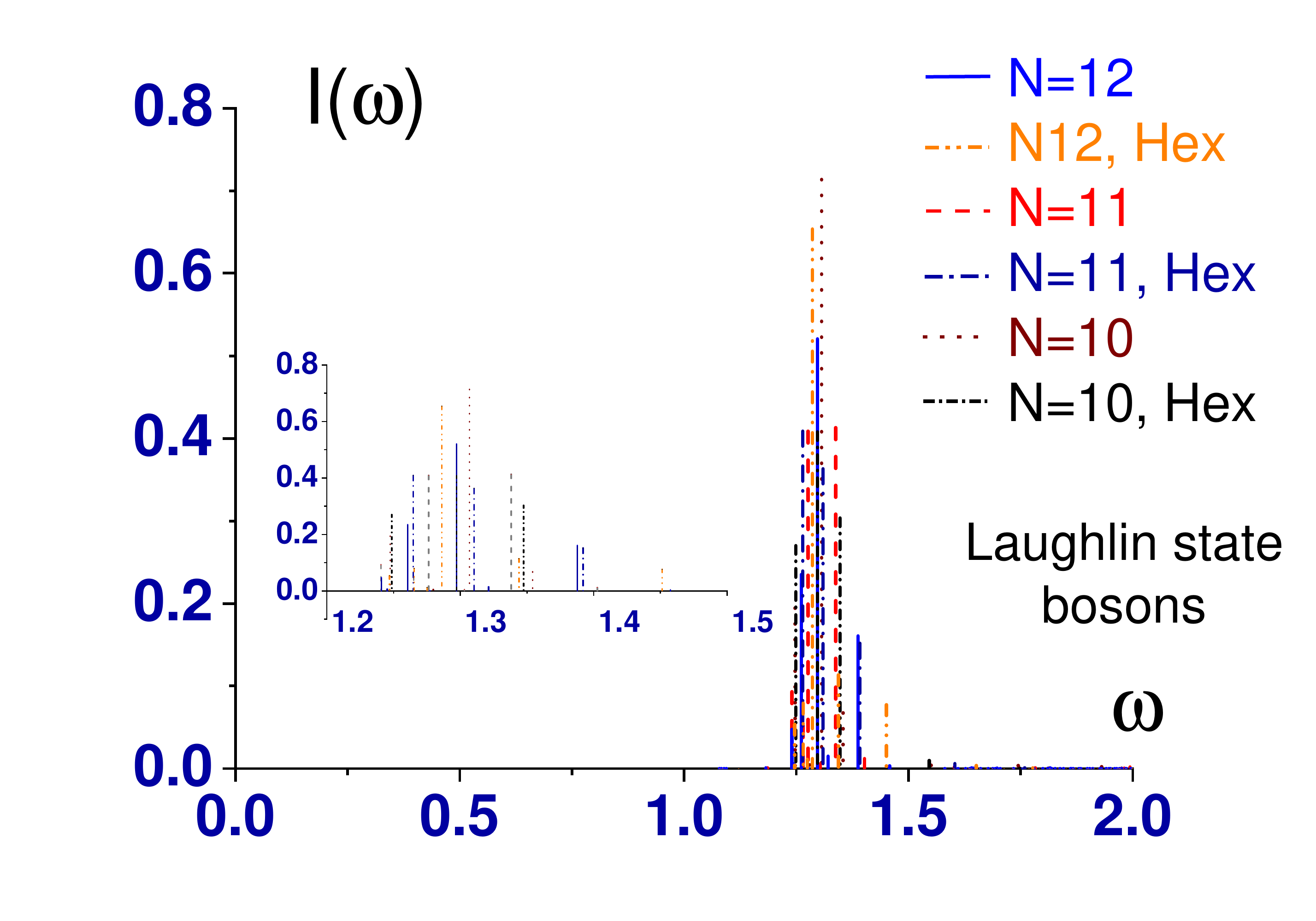}
\caption{\label{fig:LJBose}%
Same as Fig. \ref{fig:LJFermi} except for bosons at $\nu=1/2$. In this case we have collected 99\% of the total weight. The graviton response again stands out against the
background
noise. For all sizes and geometries the weights shown in the inset constitute over 98\% of the total.
}%
\end{figure}

We first turn to the Laughlin states. We consider both cases of fermions ($\nu=1/3$)
as well as bosons ($\nu=1/2$) on toroidal geometries and evaluate Eq. (\ref{eq:SpectralFunction}). We
have studied sizes up to 12 particles. The latter is generally believed to be larger than the correlation length (or size) of the system beyond which thermodynamic
behavior becomes visible. However, quantitative effects would still persists.
For small sizes
almost the
entire weight is exhausted by a single graviton peak. For larger sizes we observe broadening of the resonance and
the appearance of smaller nearby peaks. However, the integrated weight of the resonance increases linearly with size,
 a trend that generally is not followed by the
height of a single peak. For all the sizes that we considered the graviton resonance produces
the largest response in the system. Figs. \ref{fig:LJFermi} and \ref{fig:LJBose} show these cases for fermions and bosons respectively.

The Hamiltonian for both cases consists of a single pseudo-potential for relative angular momentum 1 (fermions) or 0 (bosons). Our energies are given in units of
the strength of these pseudo-potentials.
We note that the energies for which we see graviton responses are consistent with previous numerical
calculations\cite{Liu-Gromov-Papic} where the graviton is the $k=0$ energy of Girvin-MacDonald-Platzman
magneto-roton collective mode inside the excited states continuum.

\begin{figure}[t]
\centering%
\includegraphics[width=0.48\textwidth]{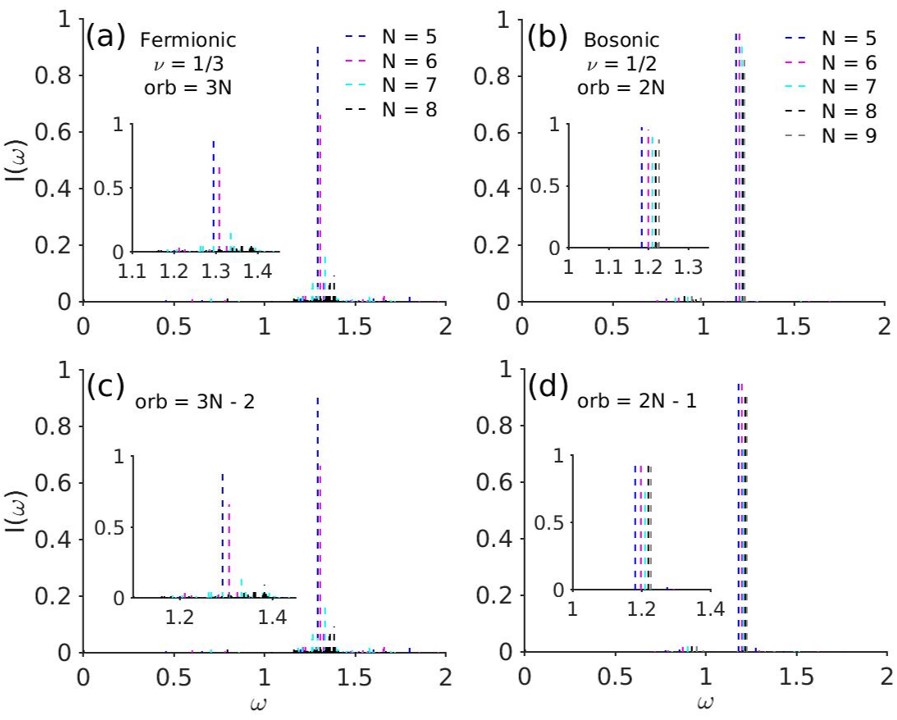}
\caption{\label{fig:LJDisk}%
A bird's-eye view of (a) and (c) fermionic $I(\omega)$ with $3N$ and $3N - 2$ orbitals and of (b) and (d) bosonic $I(\omega)$ with $2N$ and $2N - 1$ orbitals for
$N$ particles on disk geometry. For all sizes, we recovered up to 97\% of the total weight for fermionic case and up to 99\% for bosonic case. In all cases, the
graviton absorptions stand out against the background noise.
}%
\end{figure}

To establish that gravitons are chiral on the torus we employ the chiral operators $\hat{O}_\mp$.
Interestingly, $\hat{O}_-$ has the same effect as the operator of Eq. (\ref{eq:gravitycoupling}), while $\hat{O}_+$ annihilates the
model state, and so $I_+(\omega)$ is identically zero. The chiral operators, as noted, can be  generalized
to  n-body pseudo-potential Hamiltonians for the Read-Rezayi sequence. We have explicitly verified this for
the 3-body Hamiltonian of the MR state. In fact, in all our plots (except the Coulomb interaction case below)
 we show $I_-(\omega)$.

\begin{figure}[b]\centering
\includegraphics[width=0.50\textwidth]{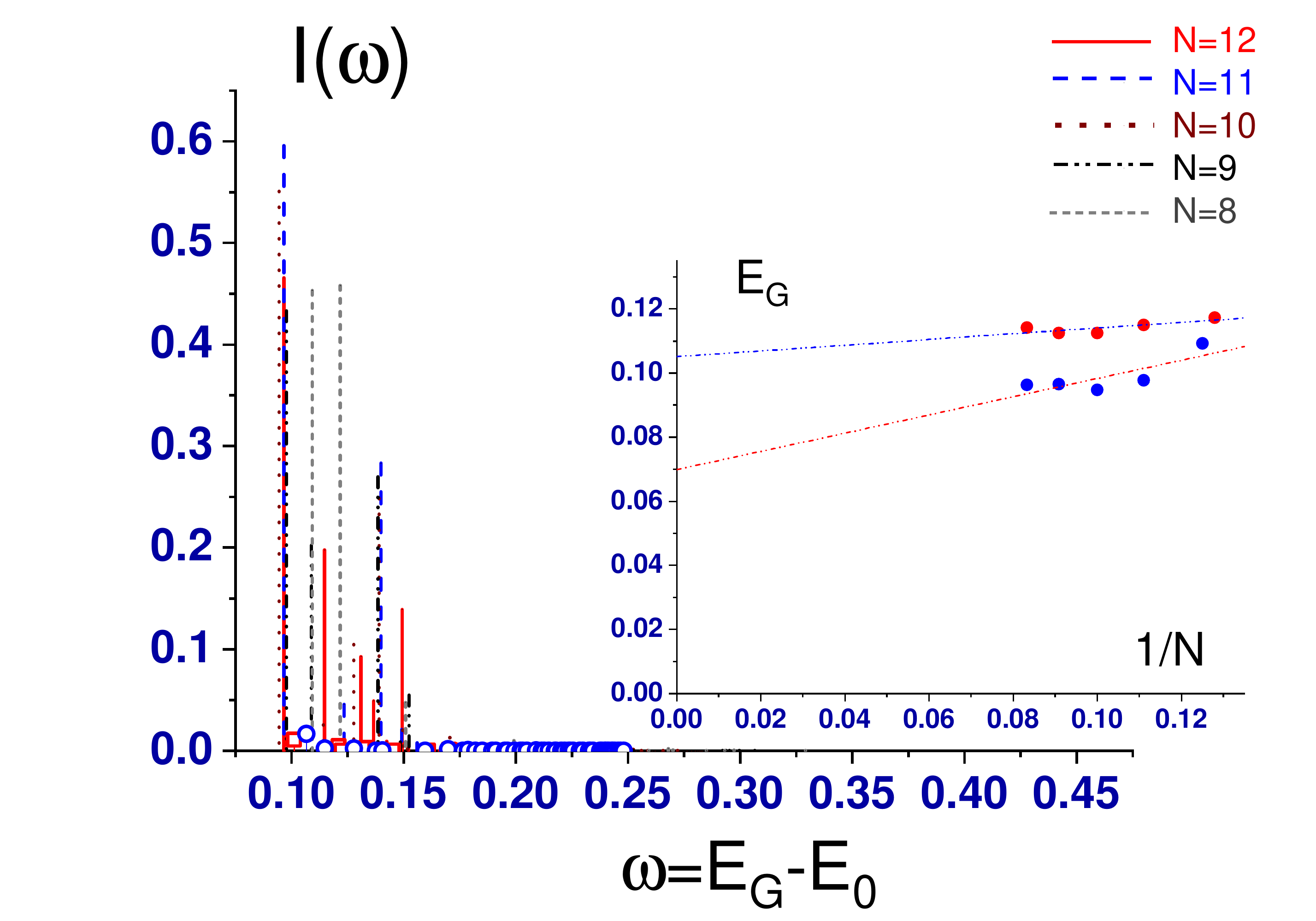}
\caption{\label{fig:Coul}
The graviton response of the Coulomb potential $\nu=1/3$ ground state with  hexagonal unit cell.
The x-axis represent the excitation energy measured from the ground state energy.  The large symbols at the
bottom  represent the relative weight of $I_+(\omega)$ ($N=12$ square and $N=11$ circular symbols). The inset
shows scaling of the graviton energy vs inverse of the system size.  The lower points are energies of the main  
peak, whereas the upper points are the average energies weighted by the size of the corresponding peaks.
}%
\end{figure}

The operator $\hat{O}_{-}$ creates excitations with angular momentum $-2$, which is also the angular momentum of the gravitons they create. To reveal this chirality more explicitly
we also investigate the two Laughlin states on disk geometry where angular momentum is a good quantum number. Instead of using Eq. (\ref{eq:gravitycoupling_new}), we express $\hat{O}_\mp^{(2)}$ in terms of anisotropic complex pseudo-potentials\cite{Bo-Yang-et-al_pseudos}:
$O ^{(2)} _{+} \propto \sum\limits_{M} |m+2, M \rangle \langle m, M|$ and $O ^{(2)} _{-} \propto
\sum\limits_{M} |m, M \rangle \langle m + 2, M|$, where $|m, M \rangle$ is a two-body state with the relative angular momentum $m$ and
center-of-mass angular momentum $M$, with
$m = 1$ for fermions and $m = 0$ for bosons. We now see why $\hat{O}_{+}$ annihilates the Laughlin state: it tries to turn a pair with relatively angular momentum $m$ into $m+2$, which does not exist in the Laughlin state.
As a result, $I _{+}(\omega)$ is zero everywhere. On the
other hand, Fig.~\ref{fig:LJDisk}(a) and (b) show strong graviton peaks in $I _{-}(\omega)$ for fermionic and bosonic cases, respectively.
Comparing to the cases on the torus, we find good agreement in peak positions, and noticeably less broadening and background noise. The results are not sensitive to the number of orbitals we keep as long as we have enough orbitals to accommodate the Laughlin states; see Fig.~\ref{fig:LJDisk}(c) and (d).

We now return to toroidal geometry and investigate the graviton contribution to the spectral functions for the Coulomb potential at $\nu=1/3$, which is the experimentally most relevant case.
Fig.  \ref{fig:Coul} shows $I_-(\omega)$ for electrons at $\nu=1/3$ on a square torus, where we use the Coulomb interaction including finite well-thickness effects appropriate for the samples of Ref. \cite{pinczuk}, whose relation with our work is discussed below. While the weights are smaller than
those of the model states, a clear signature of the graviton is discernible in comparison to other peaks
further up in the continuum.  In this case, $\hat{O}_+$ does not annihilate the Coulomb ground state, because there do exist pairs with relative angular momentum $m=1$ in the ground state. Nonetheless, the chiral nature
of the graviton is evident by the strong suppression of the (un-normalized) $I_+(\omega)$ compared to $I_-(\omega)$. This is because such pairs are rare, reflecting the Laughlin correlation.

Our bounds on the energy of the graviton (0.07 -- 0.105 in units of $e^2/4\pi\epsilon_0\ell$, $\ell$ is the magnetic
length) shown in the inset of Fig. \ref{fig:Coul}, is consistent with the 
resonance energy (0.084) found in 
the inelastic light scattering measurement of Ref. \cite{pinczuk}. In such a two-photon process (one photon absorbed and one photon emitted, also known as Raman scattering), the total angular momentum of the two photons match that of the graviton if they have the appropriate polarization; as a result gravitons can be excited\cite{golkar}.
We therefore identify the resonance of Ref. \cite{pinczuk} as due to the graviton excitation.

To obtain a more definitive prediction one may need a more systematic sampling of the resonance energies, which appear to be broadened and hence introduce the scatter we observe in the data.

\begin{figure}%
\centering%
\includegraphics[width=0.45\textwidth]{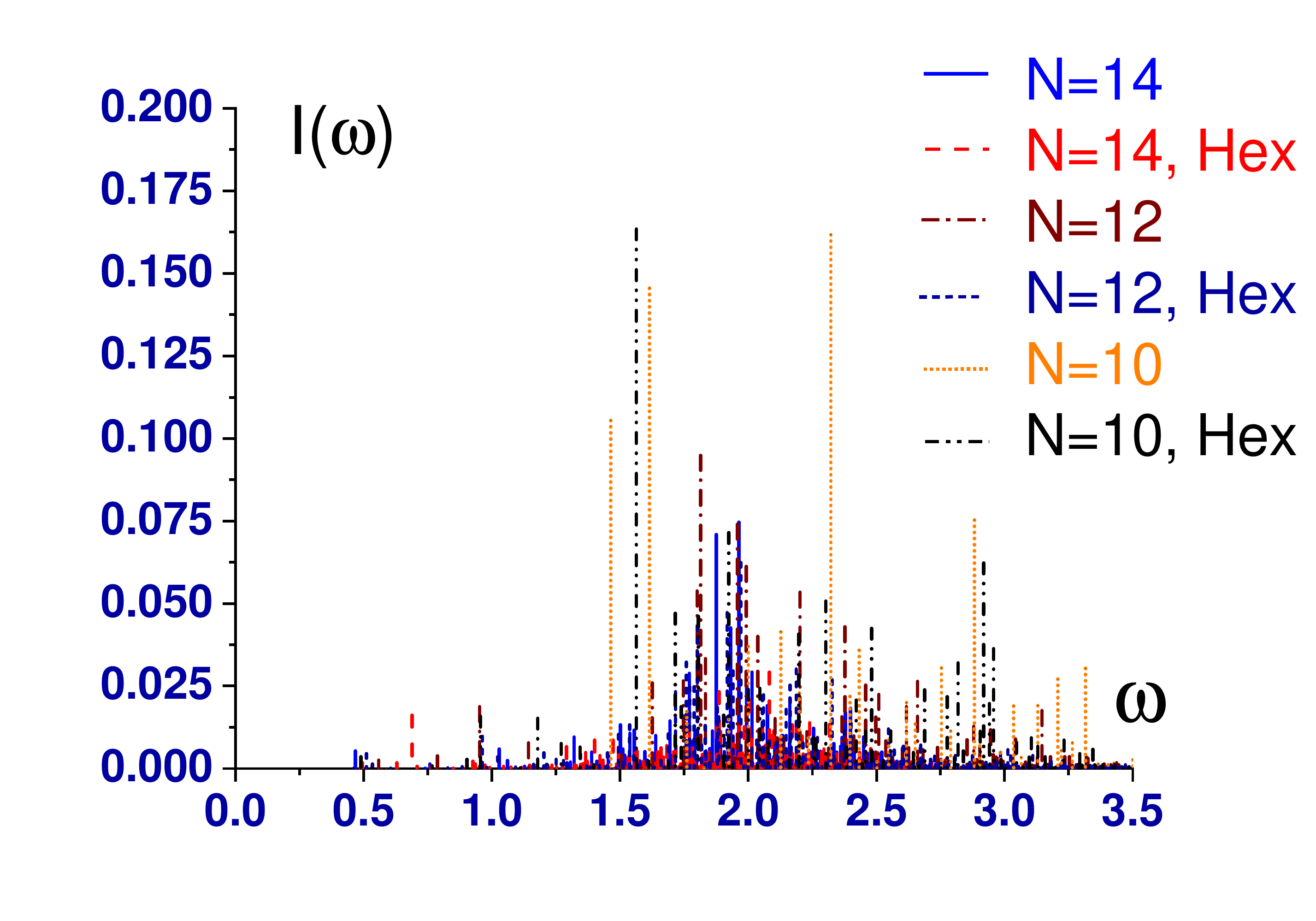}
\caption{\label{fig:MR_EV_Fermi}
The spectral function $I(\omega)$ for the MR state of fermions with even number of particles at $\nu=1/2$.
}%
\end{figure}

We next consider the MR states. As in the case of the Laughlin states,
for even numbers of particles the graviton has (orbital) spin of 2 and obeys Bose statistics. However, for
an odd number of particles, in addition to the graviton,  in Ref. \cite{boyang12} it was shown that there exists a
fermionic ``gravitino'' resonance, which  has spin-3/2.   In this paper we study the former, but
defer the investigation of gravitino to future studies.
Fig. \ref{fig:MR_EV_Fermi} shows $I_(\omega)$ (which is equivalent to $I_-(\omega)$ for the same reason as the Laughlin case) for 3-body interaction that makes the MR state exact ground state for fermions at $\nu=1/2$. For this part, we have again studied the square
and hexagonal torus, which are the two highest symmetry geometries.

In  the case of the square geometry, the 3-fold degeneracy of the MR state for an even number of electrons is split into a non-degenerate one in
the Brillouin zone (BZ) corner and a pair of degenerate states with their wave vectors on the BZ boundary. In both cases of
fermions and bosons we have included some of these ground states in our plots. In some cases for the same size and
geometry the two distinct ground states either show a single strong peak or a few
smaller neighboring peaks, but with comparable total weights, which appear to indicate broadened resonance with a shorter lifetime.  For the hexagonal geometry all
3 ground states are degenerate, and depending on size, they exhibit
both sharp and broadened peaks.

\begin{figure}%
\centering%
\includegraphics[width=0.45\textwidth]{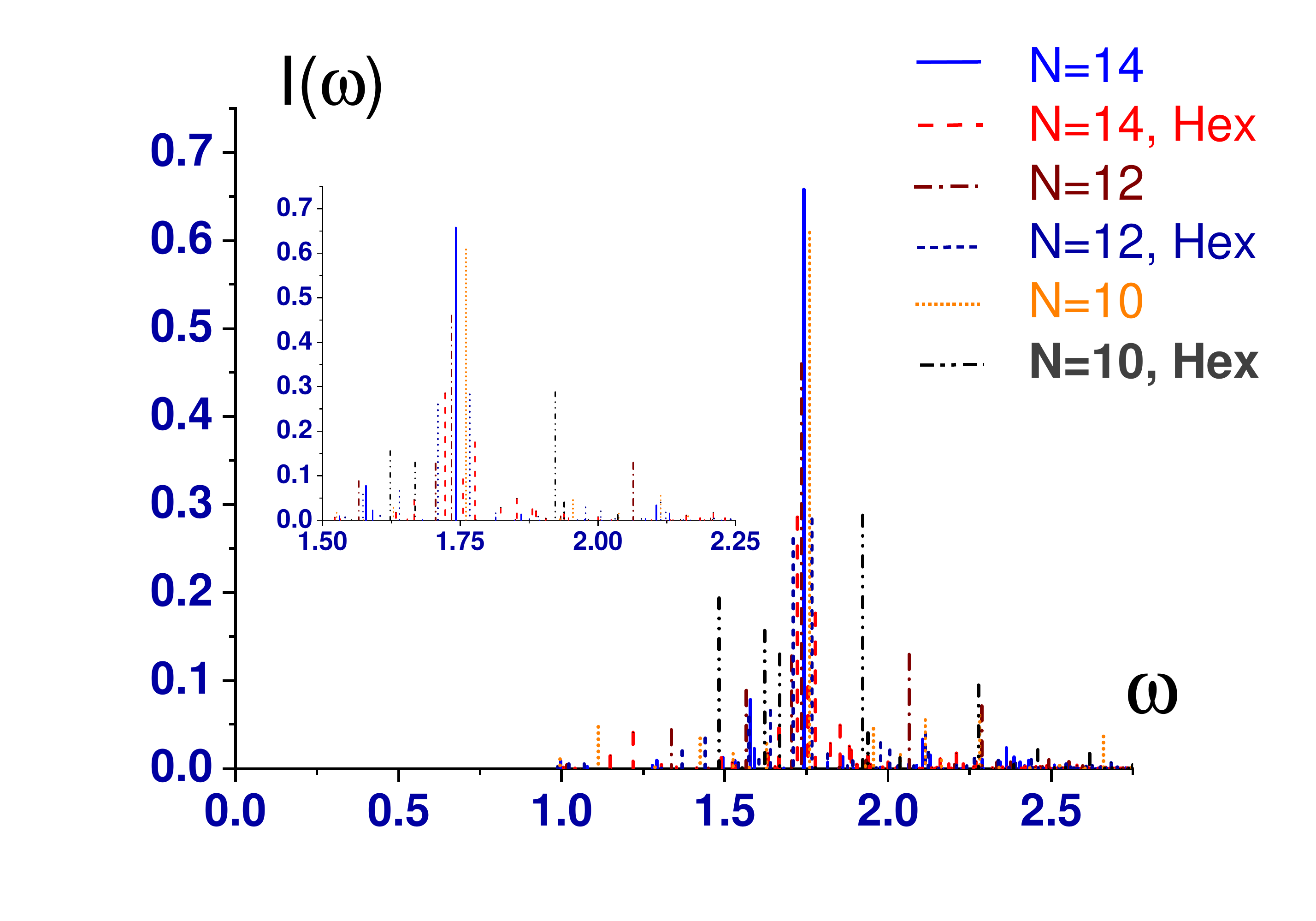}
\caption{\label{fig:MR_EV_Bose}
The spectral function $I(\omega)$ for the MR state of bosons with even number of particles at $\nu=1$.
}%
\end{figure}

\begin{figure}%
\centering%
\includegraphics[width=0.45\textwidth]{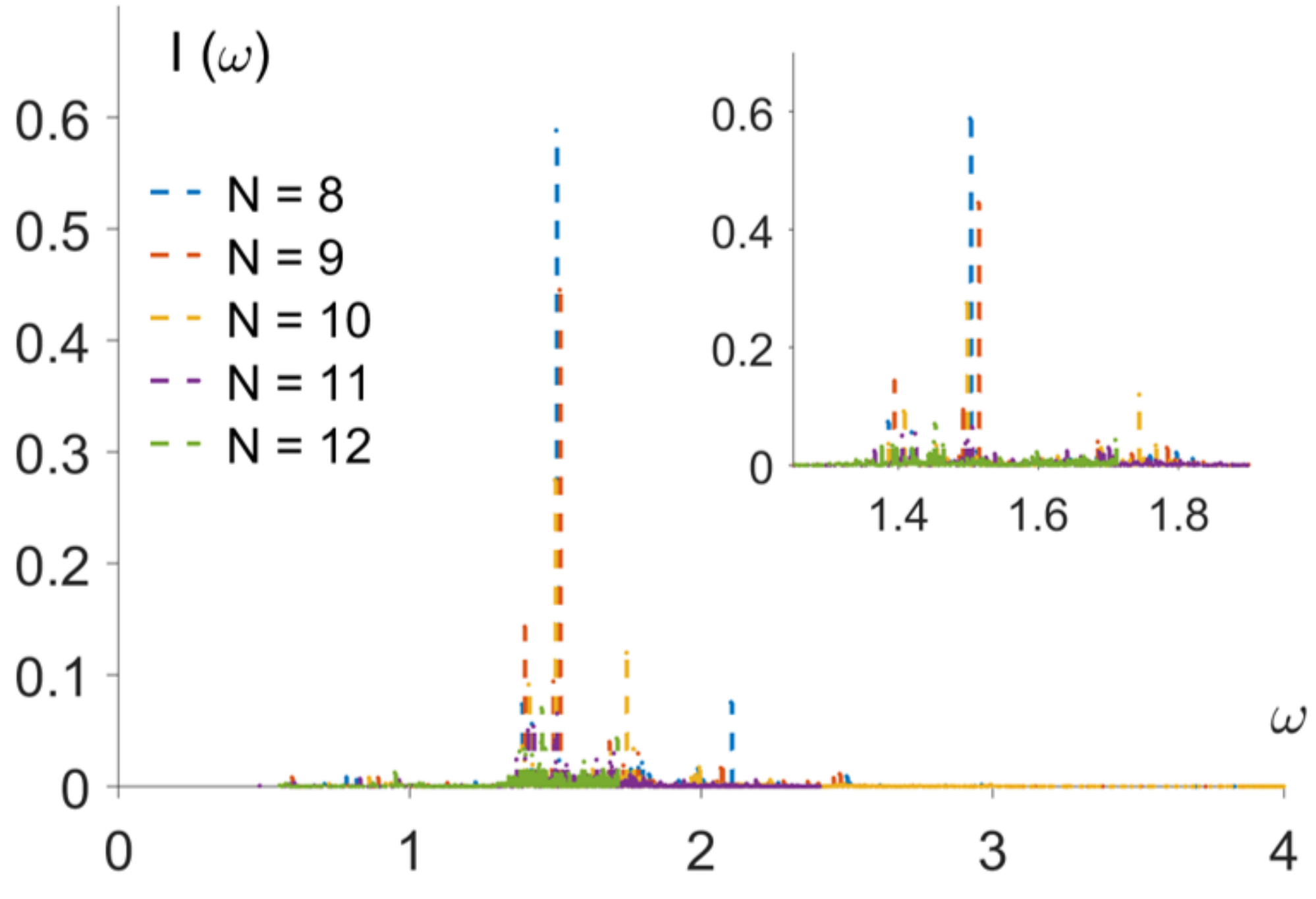}
\caption{\label{fig:MR_bosonDisc}
The spectral function $I(\omega)$ for the MR state of bosons at $\nu=1$ on a 
disc, where the orbital number used is the same as particle number $N$.
}%
\end{figure}

In Fig. \ref{fig:MR_EV_Bose} we present the results for
boson MR states at $\nu=1$. For this case we have also calculated $I_-(\omega)$ on a disc; see Fig. \ref{fig:MR_bosonDisc}. The results are consistent with those obtained on the torus, although the graviton peak appears sharper. Note that on a disc subtleties associated with the 3-fold degeneracy and differences between even and odd particle number cases do not exist, which may be contributing a factors to better quality data.

As seen by the size of the weights in these figures
the noise level increases considerably in the 3-body case and is worse for fermions.  A similar trend was
observed for the Laughlin states as noted earlier. Nevertheless, the  graviton peaks do stand out
against the background noise.

{\em Experimental Observability and Future Work} -- In Ref. \cite{yang16} it was suggested that graviton will show up as a sharp resonance peak in the absorption spectrum of acoustic wave propagating perpendicular to the 2DEG, which is an analog of the gravitational wave. Our numerical results support this, as the spectral functions calculated here are those describing the coupling between acoustic waves and 2DEG. Such experiments have yet to be performed. More recently it was argued that the graviton will dominate quench dynamics in a FQH liquid\cite{Liu-Gromov-Papic}. As discussed above we identify the resonance peak in the inelastic light scattering at $\nu=1/3$\cite{pinczuk} as due to the graviton. To gain a more detailed understanding we need to calculate the appropriate spectral function for such processes. As pointed out previously\cite{golkar} the chirality of the graviton can be revealed through the polarization of the light in a Raman process. Here we are able to make a much more specific prediction: In order to excite the chiral graviton with angular momentum $-2$, the incoming light needs to be circularly polarized to have angular momentum $-1$, while the (Raman or inelastically) scattered light will have the opposite polarization and have angular momentum $+1$, thus transferring a net angular momentum $-2$ to the 2DEG. In a very recent experimental work\cite{Du}, inelastic light scattering was performed on the first excited Landau level states. We tentatively attribute the sharp resonance at $\nu=7/3$ (which is termed a new plasmon) to the graviton, similar to that at $\nu=1/3$; the much broadened peak at $\nu=5/2$ is consistent with the broadening we found in our calculations for the MR state. But much more detailed studies using interactions appropriate to the first excited Landau level are needed.

In summary, we have found a clear signature of a chiral graviton mode for both Laughlin and MR states, and particularly
for the ground state of the Coulomb interaction at $\nu=1/3$.
In all cases of torus studies the total weights, the bulk of which constitute
the graviton resonance, scales linearly with system size. Our results are consistent with the inelastic light scattering experiment of Pinczuk et. al. that sees a resonance with zero momentum.

\begin{acknowledgments}
We thank Zlatko Papic for helpful comments.
This work was supported by DOE grant No. \protect{DE-SC0002140}.
\end{acknowledgments}

\end{document}